# Newtonian Gravitational Waves from a Continuum

PETER VADASZ

Department of Mechanical Engineering

Northern Arizona University, Flagstaff, AZ, U.S.A.

## ABSTRACT

Gravitational waves are being shown to derive directly from Newtonian dynamics for a continuous mass distribution, e.g. compressible fluids or equivalent. It is shown that the equations governing a continuous mass distribution, i.e. the inviscid Navier-Stokes equations for a general variable gravitational field $\boldsymbol{g}(t,\boldsymbol{x})$, are equivalent to a form identical to Maxwell equations from electromagnetism, subject to a specified condition. The consequence of this equivalence is the creation of gravity waves that propagate at finite speed. The latter implies that Newtonian gravitation as presented in this paper is not "spooky action at a distance" but rather is similar to electromagnetic waves propagating at finite speed, despite the apparent form appearing in the integrated field formula. In addition, this proves that in analogy to Maxwell equations the Newtonian gravitation equations are Lorentz invariant for waves propagating at the speed of light. Since gravitational waves were so far derived only from Einstein's general relativity theory it becomes appealing to check if there is a connection between the Newtonian waves presented in this paper and the general relativity type of waves at least in a certain limit of overlapping validity, i.e. as a flat-space approximation. The latter is left for a follow-up research.

## 1. Introduction

Deriving equations identical in form to Maxwell equations for continuous media has been attempted before particularly with application to fluid dynamics. For example, Marmanis [3] uses an equation derived by Lamb [1] from the incompressible Navier-Stokes equations to use it in deriving a new theory of turbulence. A similar approach was used by Sridhar [4] in order to "formulate the problem of advection and diffusion of a passive tracer by an arbitrary, incompressible velocity field", to find that the "problem is identical to the diffusive dynamics of a charged particle in electromagnetic fields constructed from the velocity field." Rousseaux *et al.* [2] tested experimentally and theoretically the concept of "hydrodynamic charge" in the case of a "coherent structure such as the Burgers vortex". These attempts apply to the incompressible fluid Navier-Stokes equations without the gravitational field and result in a form identical to Maxwell equations having the following correspondence: the electromagnetic vector potential converts into velocity, the magnetic field converts into vorticity, the electric field converts into Lamb vector ($l = -v \times \nabla \times v$), where $v(t, x)$ is the velocity, and the electric charge converts into a "hydrodynamic charge" $q_H$ identical to the divergence of the Lamb vector, i.e. $q_H = \nabla \cdot l$.

The present paper shows a different approach applied to a variable gravitational field and a continuous mass distribution similar to a compressible fluid or equivalent, leading to a form identical to Maxwell equations. The latter starts from the inviscid but compressible Navier-Stokes equations (Euler equations) and is subject to a generalized Beltrami condition expressed in the form $\nabla \times l = -\nabla \times [v \times \nabla \times v] = 0$ (Rousseaux *et al.* [2], Yoshida et al. [5], Mahajan and Yoshida [6], Gerner [7], Amari *et al.* [8], Bhattacharjee [9], Lakhatakia [10]). The derivation in the present paper shows the results emerge in a form identical to Maxwell equations, similar to Jefimenko's [11] "gravitation and cogravitation" postulate, but rather rigorously derived instead of postulating them. Therefore, one of the paper's outcomes is the emergence of a gravitation model that is analogue to the electromagnetic field theory.

There are five major conclusions and outcomes from this derivation, namely:

(1) In this model, Newtonian gravitation is not "spooky action at a distance" but rather like the electromagnetism produces waves that propagate at finite speed.

(2) These gravitational waves emerge directly from Newtonian dynamics without the need of using general relativity. (This does not imply that they are identical to the gravitational waves emerging from general relativity in a curved space-time, though they may be a flat-space approximation for the latter.)

(3) The equations governing these waves are Lorentz invariant, just like Maxwell equations, for waves propagating at the speed of light, i.e. for $v_o = c_o$.

(4) These gravitational waves in terms of $\boldsymbol{g}(t, \boldsymbol{x})$ propagate in vacuum too as demonstrated in section 4.

(5) Section 3, the Appendix, and section 5 also show a theoretical derivation of Newton's law of universal gravitation from first principles, i.e. mass conservation/balance. Although Newton's universal gravitation constant was used during the derivation, its functional form emerges as a solution rather than being used as a postulate. The origin of the functional form of Newton's law of universal gravitation is identified in the present derivation by the divergence operator in the continuity (mass conservation/balance) equation (1).

## 2. Governing Equations

Assuming a continuous mass distribution similar to a compressible fluid, the equations governing its dynamics are the continuity equation, representing conservation of mass (mass balance), and the inviscid momentum equation representing conservation of linear momentum (balance of linear momentum). The latter is the Newton's second law applied to a continuous mass medium.

$$\frac{\partial \rho}{\partial t} + \nabla \cdot (\rho \boldsymbol{v}) = 0 \tag{1}$$

$$\rho \left[ \frac{\partial \boldsymbol{v}}{\partial t} + (\boldsymbol{v} \cdot \nabla) \boldsymbol{v} \right] = -\nabla P + \rho \boldsymbol{g} \tag{2}$$

where $\rho(t, \boldsymbol{x})$ is the mass density, $\boldsymbol{v}(t, \boldsymbol{x})$ is the velocity, $P(t, \boldsymbol{x})$ is pressure, $\boldsymbol{g}(t, \boldsymbol{x})$ is the gravitational field, $t$ is time and $\boldsymbol{x}$ is the position vector. The term $\rho \boldsymbol{g}$ represents the intrinsic gravitational forces impressed between differential mass elements due to the distributed mass within the domain occupied by this mass. By using a linear approximation for the constitutive relationship between pressure and density in the form

$$P = P_o + v_o^2 (\rho - \rho_o) \tag{3}$$

where $P_o$ and $\rho_o$ are reference values of pressure and mass density, respectively, and $v_o = \sqrt{\partial P / \partial \rho} = 1/\sqrt{\rho_o \beta_P}$ is the constant speed of propagation of the pressure wave (speed of sound as a special case, but affecting the wave propagation of the gravitational field $\boldsymbol{g}$ beyond the region containing the mass, as it is shown in section 4). If the mass speed at some point in time and space reaches the value of the speed of wave propagation, i.e. $|\boldsymbol{v}| = v_o$, then as long as $v_o^2 << c_o^2$, i.e. the speed of wave propagation is much smaller than the speed of light in vacuum, one may use Galilean transformations for evaluating $v_o$ in different frames of reference, however as the value of $|\boldsymbol{v}| = v_o$ becomes closer to the speed of light in vacuum, i.e. $v_o^2/c_o^2 \sim O(1)$ the Lorentz transformations need to be used. The speed of wave propagation $v_o$ is related to the compressibility factor of the medium $\beta_P = \rho_o^{-1}(\partial \rho/\partial P)$. Therefore, this would be the most direct and simple way in evaluating the value of $v_o$. When the compressibility factor of the medium, $\beta_P$, is very small indicating an extremely dense (compact) mass, the value of $v_o = 1/\sqrt{\rho_o \beta_P}$ becomes closer to $c_o$. In section 5.2 it is shown in equation (55) that when $v_o = c_o/2$ an equilibrium solution produces the Schwarzwild radius, i.e. the event horizon as the lower limit for a star to collapse into a black hole. The approximation, eq. (3), is nothing more than a Taylor expansion of any constitutive relationship between mass density $\rho$ and pressure $P$ around the point $P_o, \rho_o$. Substituting (3) into (2) and dividing the whole equation by $\rho$ yields

$$\frac{\partial \boldsymbol{v}}{\partial t} + (\boldsymbol{v} \cdot \nabla)\boldsymbol{v} = -v_o^2 \nabla(\ln \rho) + \boldsymbol{g} \tag{4}$$

By using the following vector identity

$$(\boldsymbol{v} \cdot \nabla)\boldsymbol{v} = \nabla\left(\frac{\boldsymbol{v} \cdot \boldsymbol{v}}{2}\right) - \boldsymbol{v} \times \nabla \times \boldsymbol{v} \tag{5}$$

into (4) leads to

$$\frac{\partial \boldsymbol{v}}{\partial t} = -\nabla\left[v_o^2(\ln \rho) + \left(\frac{\boldsymbol{v} \cdot \boldsymbol{v}}{2}\right)\right] + \boldsymbol{v} \times (\nabla \times \boldsymbol{v}) + \boldsymbol{g} \tag{6}$$

## 3. Converting the Governing Equations into a Maxwell form

Taking the curl of equation (6) produces

$$\frac{\partial(\nabla \times \boldsymbol{v})}{\partial t} = \nabla \times \left[\boldsymbol{v} \times (\nabla \times \boldsymbol{v})\right] + \nabla \times \boldsymbol{g} \tag{7}$$

The curl of the velocity is related to the counter-vorticity vector $\boldsymbol{\xi}$ defined by

$$\boldsymbol{\xi} = -\nabla \times \boldsymbol{v} \tag{8}$$

It also defines the Lamb vector $\boldsymbol{l}$ in the form

$$\boldsymbol{l} = \boldsymbol{v} \times \boldsymbol{\xi} = -\boldsymbol{v} \times (\nabla \times \boldsymbol{v}) \tag{9}$$

By substituting (8) into (7) it yields

$$\nabla \times \boldsymbol{g} = -\frac{\partial \boldsymbol{\xi}}{\partial t} + \nabla \times (\boldsymbol{v} \times \boldsymbol{\xi}) \tag{10}$$

Also from (8) it follows directly that

$$\nabla \cdot \boldsymbol{\xi} = 0 \tag{11}$$

Equations (10) and (11) emerged directly from the inviscid momentum equation (2).

If the following relationships are being introduced

$$\nabla \cdot \boldsymbol{g} = -4\pi G \rho \tag{12}$$

$$v_o^2 \nabla \times \boldsymbol{\xi} = -4\pi G \rho \boldsymbol{v} + \frac{\partial \boldsymbol{g}}{\partial t} \tag{13}$$

where $G$ is the universal gravitational constant, the continuity equation (1) is then identically satisfied as can be checked by applying the divergence operator on equation (13), using the vector identity $\nabla \cdot (\nabla \times \boldsymbol{\xi}) = 0$, and substituting the result into equation (12). However, this does not yet show how equations (12) and (13) emerged directly from equation (1). This is shown rigorously in the Appendix demonstrating that equation (1) leads directly to equations (12) and (13). Consequently one can deduce that equations (12) and (13) are equivalent to the continuity equation (1) and equations (10) and (11) are equivalent to the momentum equation (2). Moreover, equations (12) and (13) emerged from equation (1) and equations (10) and (11) emerged from equation (2).

Equation (10) becomes identical to Faraday law of induction in the Maxwell equations if

$$\nabla \times (\boldsymbol{v} \times \boldsymbol{\xi}) = 0 \tag{14}$$

Condition (14) when the curl of the Lamb vector is zero is known as the generalized Beltrami condition producing generalized Beltrami flows. It is satisfied in the following special cases

(a) Irrotational flow when $\boldsymbol{\xi} = -\nabla \times \boldsymbol{v} = 0$, that introduces a potential $V$, such that $\boldsymbol{v} = -\nabla V$, hence this condition produces potential flow,

(b) Beltrami flow when the Lamb vector is zero, i.e. $\boldsymbol{l} = \boldsymbol{v} \times \boldsymbol{\xi} = 0$ and the counter-vorticity is colinear with the velocity (they are parallel to each other), i.e. $\boldsymbol{\xi} = \alpha(t, \boldsymbol{x}) \boldsymbol{v}$. This type of flow cannot be two dimensional.

(c) Generalized Beltrami flow when $\nabla \times (\boldsymbol{v} \times \boldsymbol{\xi}) = 0$, implying that $\boldsymbol{v} \times \boldsymbol{\xi} = \nabla U$, i.e. $(\boldsymbol{v} \times \boldsymbol{\xi})$ has a potential $U$.

Subject to condition (14) equation (10) becomes

$$\nabla \times \boldsymbol{g} = -\frac{\partial \boldsymbol{\xi}}{\partial t} \qquad (15)$$

Equations (11), (12), (13) and (15) are identical to Maxwell equations in free space with the analogy as presented in Table 1. They represent the Newtonian gravitational field equations.

Having demonstrated the equivalence between the inviscid Navier-Stokes equations subject to the generalized Beltrami condition (14) and the form identical to Maxwell equations one can reach an additional conclusion. Since Maxwell equations are Lorentz invariant it implies that the inviscid Navier-Stokes equations subject to condition (14) are also Lorentz invariant for waves propagating at the speed of light, i.e. for $v_o = c_o$.

Table 1: Correspondence of variables and parameters from the inviscid Navier-Stokes to Maxwell equations

| Fluid Flow (inviscid Navier-Stokes equations) | Electromagnetism (Maxwell equations) |
| :---: | :---: |
| $\rho$ | $\rho_q$ |
| $\boldsymbol{g}$ | $\boldsymbol{E}$ |
| $\boldsymbol{\xi}$ | $\boldsymbol{B}$ |
| $-1/4\pi G$ | $\varepsilon_o$ |
| $v_o^2$ | $c_o^2$ |

where $\rho_q$ is the charge density, $\boldsymbol{E}$ is the electric field, $\boldsymbol{B}$ is the magnetic field, $\varepsilon_o$ is the permittivity of vacuum, $c_o$ is the speed of light in vacuum, and $\boldsymbol{J_q} = \rho_q \boldsymbol{v}$ is the electric current density (charge flux).

## 4. Wave Equations due to the Newtonian Gravitational Field

Substituting (8) into (15) leads to

$$\nabla \times \boldsymbol{g} = \nabla \times \left( \frac{\partial \boldsymbol{v}}{\partial t} \right) \qquad (16)$$

which can be presented in the form

$$\nabla \times \left[ \boldsymbol{g} - \frac{\partial \boldsymbol{v}}{\partial t} \right] = 0 \qquad (17)$$

It becomes appealing to define a potential $\psi$ such that

$$\boldsymbol{g} - \frac{\partial \boldsymbol{v}}{\partial t} = \nabla \psi \tag{18}$$

for any arbitrary function $\psi$, leading to

$$\boldsymbol{g} = \frac{\partial \boldsymbol{v}}{\partial t} + \nabla \psi \tag{19}$$

Substituting (19) into (12) produces

$$\frac{\partial}{\partial t}(\nabla \cdot \boldsymbol{v}) + \nabla^2 \psi = -4\pi G \rho \tag{20}$$

By using the Lorenz gauge relation converted to this gravitational problem in the form (since $\psi$ is arbitrary, one is free to choose it in a way that follows the Lorenz gauge relation)

$$\nabla \cdot \boldsymbol{v} = -\frac{1}{v_o^2} \frac{\partial \psi}{\partial t} \tag{21}$$

and substituting (21) into (20) yields

$$\frac{1}{v_o^2} \frac{\partial^2 \psi}{\partial t^2} = \nabla^2 \psi + 4\pi G \rho \tag{22}$$

Substituting (8) and (19) into (13) leads to

$$-v_o^2 \nabla \times \nabla \times \boldsymbol{v} = -4\pi G \rho \boldsymbol{v} + \frac{\partial^2 \boldsymbol{v}}{\partial t^2} + \nabla \left( \frac{\partial \psi}{\partial t} \right) \tag{23}$$

and substituting the vector identity $\nabla \times \nabla \times \boldsymbol{v} = \nabla(\nabla \cdot \boldsymbol{v}) - \nabla^2 \boldsymbol{v}$ and the Lorenz gauge relation (21) into (23) produces

$$\frac{1}{v_o^2} \frac{\partial^2 \boldsymbol{v}}{\partial t^2} = \nabla^2 \boldsymbol{v} + \frac{4\pi G}{v_o^2} \rho \boldsymbol{v} \tag{24}$$

Equations (22) and (24) are the gravitational wave equations that emerged from Newtonian dynamics for continuous media. For any given mass density distribution $\rho(\boldsymbol{x})$ the solution to (24) produces the velocity wave $\boldsymbol{v}(t, \boldsymbol{x})$ and the solution to (22) produces the function $\psi(t, \boldsymbol{x})$. Substituting these solutions of $\boldsymbol{v}(t, \boldsymbol{x})$ and $\psi(t, \boldsymbol{x})$ into (19) leads to the gravitational field solution $\boldsymbol{g}(t, \boldsymbol{x})$. The Newtonian gravitational wave equations (22) and (24) produce waves that propagate at finite speeds. Consequently, Newton's gravitation is not "spooky action at a distance" despite the apparent form appearing in its integral field formula.

There is no meaning to velocity without the mass related to it. Therefore, the waves in terms of $v(t, \boldsymbol{x})$ do not extend beyond the spherical domain containing the mass, but only within $r \leq r_o$. However, even for $\boldsymbol{v} = 0$ substituted into (19) $\boldsymbol{g} = \nabla \psi$ and the gravitational wave $\boldsymbol{g}(t, \boldsymbol{x})$ can and does propagate beyond the domain containing the mass, i.e. for $r > r_o$ too, as presented in what follows. In regions away from the distributed mass (where $r > r_o$) the pressure $P = 0$, and velocity is zero $\boldsymbol{v} = 0$ not because the mass is at rest, but because there is no mass (vacuum) and $\rho = 0$ too. Then, the wave equation (22) becomes

$$\frac{1}{v_o^2} \frac{\partial^2 \psi}{\partial t^2} = \nabla^2 \psi \tag{25}$$

and equation (19) becomes

$$\boldsymbol{g} = \nabla \psi \tag{26}$$

Consequently, equation (25) being still a wave equation (actually the classical wave equation) the solutions for $\psi(t, \boldsymbol{x})$ and based on (26) also for $\boldsymbol{g}(t, \boldsymbol{x})$ are still wave solutions. The value of $v_o$ is established in the region creating the field, where mass is present, but the resulting wave equation (25) and its link to the gravitational field (26) are clear evidence that these waves do propagate beyond the region of the mass distribution creating the field. The Newtonian gravitational waves presented in this paper do propagate in vacuum, as just demonstrated.

## 5. Specific Solutions to Newtonian Gravitational Field Equations

The solutions to the Newtonian gravitational field equations can be obtained from equations (12), (13), (15), and (11) presented in the form

$$\frac{1}{4\pi G} \nabla \cdot \boldsymbol{g} = -\rho \tag{27}$$

$$\frac{\partial \boldsymbol{g}}{\partial t} = v_o^2 \nabla \times \boldsymbol{\xi} + 4\pi G \rho \boldsymbol{v} \tag{28}$$

$$\frac{\partial \boldsymbol{\xi}}{\partial t} = -\nabla \times \boldsymbol{g} \tag{29}$$

$$\nabla \cdot \boldsymbol{\xi} = 0 \tag{30}$$

Equation (27) that was one of the two equations obtained directly from the continuity equation (1) can produce the familiar Poisson equation for Newtonian gravitation that is known to be a special case of Einstein's theory of general relativity. By assuming that the gravitational field is conservative, i.e. $\boldsymbol{g} = -\nabla\phi$ and substituting the latter into (27) produces the Poisson equation for Newtonian gravitation in the form $\nabla^2\phi = 4\pi G\rho$. However, as distinct from the standard derivation of the Poisson equation for Newtonian gravitation that includes deriving equation (27) by starting explicitly from the algebraic form of the Newton's law of universal gravitation, i.e. $\boldsymbol{g} = -\left(Gm_o/r^2\right)\hat{\boldsymbol{e}}_r$ the present derivation did not use the latter in deriving equation (27) that leads to the Poisson equation for Newtonian gravitation. On the contrary, Newton's law of universal gravitation is obtained as the solution instead of postulating it upfront.

## 5.1 – *Gravitational field due to a specified mass density distribution*

The solution to equation (27) for the gravitational field outside a spherical domain containing a specified distributed mass assuming spherical symmetry, i.e. $r \in \left[0, r_o\right]$ such that

$$\rho = \rho_o(r) \ \ \forall r \in \left[0, r_o\right] \text{ and } \rho = 0 \ \ \forall r > r_o \tag{31}$$

can be obtained by solving the resulting ordinary differential equation that emerges from (27) by using the spherical symmetry assumption $g_\theta = g_\varphi = 0$, $\partial(\cdot)/\partial\theta = \partial^2(\cdot)/\partial\theta^2 = 0$, and $\partial(\cdot)/\partial\varphi = \partial^2(\cdot)/\partial\varphi^2 = 0$ in the form

$$\frac{1}{r^2}\frac{\mathrm{d}\left(r^2 g_r\right)}{\mathrm{d}r} = -4\pi G\rho \tag{32}$$

where $\boldsymbol{g} = g_r(r)\hat{\boldsymbol{e}}_r$, and $\hat{\boldsymbol{e}}_r$ is a unit vector in the radial direction. Equation (32) leads to

$$\mathrm{d}\left(r^2 g_r\right) = -4\pi G\rho r^2\,\mathrm{d}r \tag{33}$$

which upon integration between $r = 0$ and any value of $r > r_o$ produces by using (31)

$$r^2 g_r(r) = -4\pi G \int_0^{r_o} \rho r^2\,\mathrm{d}r \ ; \quad \forall r > r_o \tag{34}$$

However the mass $m_o$ confined within the spherical domain is

$$m_o = 4\pi \int_0^{r_o} \rho r^2\,\mathrm{d}r \tag{35}$$

and therefore substituting (35) into (34) yields the familiar Newton's law of universal gravitation in terms of the gravitational field

$$g_r = -G\frac{m_o}{r^2} \ ; \quad \forall r > r_o \quad . \tag{36}$$

This solution applies to any radial density distribution $\rho_o(r)$ in (31). The gravitational field obtained and presented in equation (36) extends beyond the spherical region containing the mass creating the field. It is this form that created the idea of apparent "spooky action at a distance".

The solution to equation (27) for the gravitational field inside a spherical domain containing a specified distributed mass assuming spherical symmetry consistent with (31) is obtained from integrating (33) between 0 and $r \leq r_o$ leading to

$$r^2 g_r(r) = -4\pi G \int_0^r \rho r^2 \, \mathrm{d}r \quad ; \qquad \forall \, r \leq r_o \tag{37}$$

For a uniform density distribution $\rho_o = \mathrm{const.}$, the integral (37) yields

$$r^2 g_r(r) = -\frac{4\pi r^3}{3} \rho_o G \quad ; \qquad \forall \, r \leq r_o \tag{38}$$

The term $\rho_o\left(4\pi r^3/3\right)$ is the mass within the sphere of radius $r$. It produces the familiar gravitational field

$$g_r = -\frac{4\pi}{3} \rho_o G r \quad ; \qquad \forall \, r \leq r_o \tag{39}$$

that leads to simple harmonic motion when impressed on any test mass. This solution applies only to a constant (uniform) density distribution, i.e. $\rho_o = \mathrm{const.}$ in (31).

## 5.2 – Gravitational field and the solution for mass density distribution

The solutions for the gravitational field presented so far do not lead to the solution for the corresponding mass density distribution, i.e. $\rho(t, \boldsymbol{x})$. In all cases the gravitational field was obtained in terms of a specific mass density distribution. The objective of this section is to derive also the solution to the mass density distribution within a finite spherical domain. The solution inside the spherical domain containing the distributed mass assuming spherical symmetry, i.e. $r \in [0, r_o]$ such that equation (31) still applies, is due to the intrinsic gravitational forces impressed between differential mass elements due to the distributed mass. For this case there is a possible equilibrium mass density distribution and the corresponding equilibrium gravitational field distribution that can be obtained from (27), (28) and (2) subject to $\boldsymbol{v} = 0$. Substituting $\boldsymbol{v} = 0$, and equations (3), (27) into (2) yields

$$v_o^2 \nabla(\nabla \cdot \boldsymbol{g}) - \boldsymbol{g}(\nabla \cdot \boldsymbol{g}) = 0 \tag{40}$$

Assuming spherical symmetry this equation becomes

$$\frac{d}{dr}\left[\frac{1}{r^2}\frac{d\left(r^2 g_r\right)}{dr}\right]-\frac{g_r}{v_o^2 r^2}\frac{d\left(r^2 g_r\right)}{dr}=0 \tag{41}$$

where $\boldsymbol{g}=g_r\left(r\right)\hat{e}_r$. Introducing the notation

$$\eta=r^2 g_r \tag{42}$$

into (41) produces the equation

$$\frac{d}{dr}\left[\frac{1}{r^2}\frac{d\eta}{dr}\right]-\frac{\eta}{v_o^2 r^4}\frac{d\eta}{dr}=0 \tag{43}$$

which can be presented in the expanded form

$$\frac{1}{r^2}\frac{d^2\eta}{dr^2}-\left(2+\frac{\eta}{v_o^2 r}\right)\frac{1}{r^3}\frac{d\eta}{dr}=0 \tag{44}$$

Equation (44) has a possible solution by setting the term in the brackets equal to zero, i.e.

$$2+\frac{\eta}{v_o^2 r}=0 \tag{45}$$

leading to

$$\eta=-2v_o^2 r \tag{46}$$

The solution (46) satisfies also $d^2\eta/dr^2=0$ and consequently satisfying the whole equation (44). Reverting back to the original variable by using (42) into (46) leads to the solution for the gravitational field

$$g_r=-\frac{2v_o^2}{r} \tag{47}$$

or

$$\boldsymbol{g}=-\frac{2v_o^2}{r}\hat{e}_r \tag{48}$$

showing that the gravitational field is attracting as identified by the negative sign. Substituting this solution into equation (27) expressed for spherical symmetry conditions, i.e.

$$\rho=-\frac{1}{4\pi G r^2}\frac{d\left(r^2 g_r\right)}{dr} \tag{49}$$

yields for the mass density distribution the equilibrium solution

$$\rho=\frac{v_o^2}{2\pi G}\frac{1}{r^2} \tag{50}$$

Since the space occupied by the continuously distributed mass consists of a total mass $m_o$ it implies

$$\int\limits_0^{\tilde{V}_o} \rho\, \mathrm{d}\tilde{V} = m_o \tag{51}$$

where $\mathrm{d}\tilde{V} = r^2 \sin\theta\, \mathrm{d}r\, \mathrm{d}\theta\, \mathrm{d}\varphi$ is the differential volume, $\varphi \in \left[0, 2\pi\right]$, $\theta \in \left[0, \pi\right]$ are the azimuthal and polar coordinates, respectively, while $\tilde{V}_o$ is the total volume containing the mass $m_o$. Then, in spherical coordinates

$$\int\limits_0^{2\pi} \mathrm{d}\varphi \int\limits_0^{\pi} \sin\theta\, \mathrm{d}\theta \int\limits_0^{r_o} \rho r^2\, \mathrm{d}r = m_o \tag{52}$$

leading to

$$\int\limits_0^{r_o} \rho r^2\, \mathrm{d}r = \frac{m_o}{4\pi} \tag{53}$$

Substituting the solution (50) into (53) and integrating yields

$$v_o^2 = \frac{m_o G}{2 r_o} \tag{54}$$

or if expressed in terms of the radius $r_o$ versus $v_o^2$ it produces for $v_o = c_o/2$ the Schwarzschild radius, $r_S$, i.e. the lower limit for a star to collapse into a black hole, i.e.

$$r_o = \frac{2 m_o G}{c_o^2} = r_S \tag{55}$$

Equation (54) produced a relationship between the speed of wave propagation and the radius confining the distributed mass. It allows presenting the gravitational field as well as the mass density distribution solutions (48) and (50) in terms of $r_o$ in the form

$$\mathbf{g} = -\frac{m_o G}{r_o r}\hat{\mathbf{e}}_r \tag{56}$$

$$\rho = \frac{m_o}{4\pi r_o}\frac{1}{r^2} \tag{57}$$

This equilibrium solution might or might not be stable. For investigating the stability of this equilibrium one needs to consider the equations when $v \neq 0$. Preliminary results from a linear stability analysis of this equilibrium indicate that the basic equilibrium solution might be neutrally stable with oscillations subject to some conditions. The details are left for a separate presentation.

## 6. Newtonian Gravitational Waves not Subject to the Generalized Beltrami Condition

It should be also noted that even without imposing the generalized Beltrami condition, i.e. using equation (10) instead of (15) the result is still producing gravitational waves, although different than the electromagnetic ones. The latter is demonstrated in what follows. Following the same procedure as in section 4 but using equation (10) instead of equation (15), i.e. the corresponding equation obtained without using the generalized Beltrami assumption (14), and defining the potential $\psi$ in the form that is consistent with

$$\boldsymbol{g} = \frac{\partial \boldsymbol{v}}{\partial t} + \nabla \psi - \boldsymbol{v} \times (\nabla \times \boldsymbol{v}) \tag{58}$$

instead of (19), one obtains the following equations replacing equations (22) and (24), respectively

$$\frac{1}{v_o^2} \frac{\partial^2 \psi}{\partial t^2} = \nabla^2 \psi + 4\pi G \rho - \nabla \bullet \left[ \boldsymbol{v} \times (\nabla \times \boldsymbol{v}) \right] \tag{59}$$

$$\frac{1}{v_o^2} \frac{\partial^2 \boldsymbol{v}}{\partial t^2} = \nabla^2 \boldsymbol{v} + \frac{4\pi G}{v_o^2} \rho \boldsymbol{v} + \frac{1}{v_o^2} \frac{\partial}{\partial t} \left[ \boldsymbol{v} \times (\nabla \times \boldsymbol{v}) \right] \tag{60}$$

The gravitational waves emerging as the solution from these equations are distinct from the ones presented in section 4 that comply with the generalized Beltrami assumption (14), but are nevertheless still gravitational waves. Furthermore, in regions away from the distributed mass (where $r > r_o$) where the pressure $P = 0$, the velocity $\boldsymbol{v} = 0$ because there is no mass (vacuum), and $\rho = 0$ the resulting wave equations are identical to the ones obtained when using the generalized Beltrami assumption (14), as then equations (58) and (59) are identical to equations (26) and (25), respectively. The only difference being the boundary conditions at $r = r_o$, which are a result of the wave solution inside the domain containing the mass creating the field, i.e. for $r < r_o$.

The one important impact of not imposing the generalized Beltrami condition is that one still needs to prove that the resulting equations are Lorentz invariant. Lorentz invariance does apply to Maxwell equations but it is not yet proven for the extended ones, discussed in this section.

## 7. Conclusions

Newtonian gravitational waves were shown to emerge directly from Newtonian dynamics for a continuous mass distribution. The equations governing a continuous mass distribution for a general variable gravitational field $\boldsymbol{g}(t, \boldsymbol{x})$ were proven equivalent to a form identical to Maxwell equations from electromagnetism, subject to a generalized Beltrami condition. The consequence of this equivalence is the creation of gravity waves that propagate at finite speed. Consequently the latter indicates that Newtonian gravitation as presented in this paper is not "spooky action at a distance" but rather is similar to electromagnetic waves propagating at finite speed, despite the apparent form appearing in the integrated field formula. This result also proves that in analogy to Maxwell equations the Newtonian gravitation equations (derived from the inviscid Navier-Stokes equations) are Lorentz invariant for waves propagating at the speed of light, i.e. $v_o = c_o$. The connection between gravitational waves derived from Einstein's general relativity theory and the Newtonian waves presented in this paper at least in a certain limit of overlapping validity, i.e. as a flat-space approximation is a topic of further interest. The latter is left for a follow-up research.

## Appendix

In section 3 it was shown that equations (12) and (13) lead directly to the continuity equation (1) and consequently it was argued that as a result these equations are equivalent to the equation (1). In this appendix the opposite direction is demonstrated, i.e. that starting with equation (1) one obtains directly equations (12) and (13) and therefore the latter are a result and consequence of the continuity equation. Starting from equation (1)

$$\frac{\partial \rho}{\partial t} + \nabla \cdot (\rho \boldsymbol{v}) = 0 \tag{A-1}$$

it is evident that in order for both terms in the equation to share the same operator one possible way is to have a relationship between the mass density $\rho$ and (at this stage) an arbitrary vector field $\boldsymbol{f_1}$ in the form

$$\rho = C_1 \nabla \bullet \boldsymbol{f_1} \tag{A-2}$$

where $C_1$ is an arbitrary constant. Then equation (A-1) becomes

$$\nabla \bullet \left[ C_1 \frac{\partial \boldsymbol{f_1}}{\partial t} + \rho \boldsymbol{v} \right] = 0 \tag{A-3}$$

Equation (A-3) is identically satisfied if the term inside the brackets is proportional to the curl of another vector field, e.g.

$$C_1 \frac{\partial \boldsymbol{f_1}}{\partial t} + \rho \boldsymbol{v} = C_2 \nabla \times \boldsymbol{\xi} \tag{A-4}$$

Dividing equation (A-4) by $C_1$ leads to

$$\frac{\partial \boldsymbol{f_1}}{\partial t} + \frac{1}{C_1} \rho \boldsymbol{v} = C_3 \nabla \times \boldsymbol{\xi} \tag{A-5}$$

where $C_3 = C_2 / C_1$. Choosing now to identify the vector field $\boldsymbol{f_1}$ to be the gravitational field $\boldsymbol{g}$, i.e.

$$\boldsymbol{f_1} = \boldsymbol{g} \tag{A-6}$$

the second vector field $\boldsymbol{\xi}$ to be identical to the counter-vorticity, choosing the constants $C_1$ and $C_3$ in the form

$$C_1 = -\frac{1}{4\pi G} \quad \text{and} \quad C_3 = v_o^2 \tag{A-7}$$

and substituting (A-6) and (A-7) into equation (A-5) yields

$$\frac{\partial \boldsymbol{g}}{\partial t} - 4\pi G \rho \boldsymbol{v} = v_o^2 \nabla \times \boldsymbol{\xi} \tag{A-8}$$

which is identical to Ampere law equation (13).

Returning to equation (A-1), another possible way to get both terms to share the same operator is by imposing a linear relationship between the term $\rho \boldsymbol{v}$ and the time derivative of another vector field $\boldsymbol{f_2}$ in the form

$$\rho \boldsymbol{v} = C_4 \frac{\partial \boldsymbol{f_2}}{\partial t} \tag{A-9}$$

Then, by substituting (A-9) into equation (A-1) the latter becomes

$$\frac{\partial}{\partial t} \left[ \rho + C_4 \nabla \bullet \boldsymbol{f_2} \right] = 0 \tag{A-10}$$

Equation (A-10) can be satisfied if the terms in the brackets are equal to any function of space but not of time, e.g. $F(\boldsymbol{x})$. Since the choice is arbitrary the simplest form is $F(\boldsymbol{x}) = \text{constant} = 0$ leading to

$$\rho + C_4 \nabla \bullet \boldsymbol{f_2} = 0 \tag{A-11}$$

which satisfies equation (A-10). Choosing now to identify the vector field $\boldsymbol{f_2}$ with the gravitational field $\boldsymbol{g}$, i.e.

$$f_2 = g \tag{A-12}$$

and choosing the constant $C_4$ in the form

$$C_4 = \frac{1}{4\pi G} \tag{A-13}$$

leads when substituted into equation (A-11) to

$$\nabla \bullet g = -4\pi G \rho \tag{A-14}$$

Equation (A-14) is identical to the Coulomb law in field form presented as equation (12) in section 3.

These derivations showed that starting with the continuity equation (1) or (A-1) leads directly to equations (12) and (13) without the need of using the algebraic functional form of Newton's law of universal gravitation. While, several arbitrary choices were made in these derivations, none of them included explicitly the algebraic functional form of Newton's law of universal gravitation. Still the choice of the coefficients $C_1$ and $C_4$ that include Newton's universal gravitational constant prevents regarding these derivations as totally independent from the experimentally established Newton's law of universal gravitation.